\documentclass[aps,prl,reprint,showpacs,superscriptaddress]{revtex4-1}
\usepackage{graphicx}
\usepackage[utf8]{inputenc}
\usepackage[english]{babel}
\usepackage{color}
\usepackage{lipsum}

\begin{document}


\title{Spin precession and spin Hall effect in monolayer graphene/Pt nanostructures}



\author{W. Savero Torres}
\affiliation{Catalan Institute of Nanoscience and Nanotechnology (ICN2), CSIC and The Barcelona Institute of Science and Technology (BIST), Campus UAB, Bellaterra, 08193 Barcelona, Spain.}

\author{J.F. Sierra}
\affiliation{Catalan Institute of Nanoscience and Nanotechnology (ICN2), CSIC and The Barcelona Institute of Science and Technology (BIST), Campus UAB, Bellaterra, 08193 Barcelona, Spain.}

\author{L.A. Ben\'{i}tez}
\affiliation{Catalan Institute of Nanoscience and Nanotechnology (ICN2), CSIC and The Barcelona Institute of Science and Technology (BIST), Campus UAB, Bellaterra, 08193 Barcelona, Spain.}
\affiliation{Universidad Autonoma de Barcelona, Bellaterra, 08010 Barcelona, Spain}

\author{F. Bonell}
\affiliation{Catalan Institute of Nanoscience and Nanotechnology (ICN2), CSIC and The Barcelona Institute of Science and Technology (BIST), Campus UAB, Bellaterra, 08193 Barcelona, Spain.}

\author{M.V. Costache}
\affiliation{Catalan Institute of Nanoscience and Nanotechnology (ICN2), CSIC and The Barcelona Institute of Science and Technology (BIST), Campus UAB, Bellaterra, 08193 Barcelona, Spain.}

\author{S.O. Valenzuela}
\affiliation{Catalan Institute of Nanoscience and Nanotechnology (ICN2), CSIC and The Barcelona Institute of Science and Technology (BIST), Campus UAB, Bellaterra, 08193 Barcelona, Spain.}
\affiliation{Instituci\'{o} Catalana de Recerca i Estudis Avan\c{c}ats (ICREA), 08070 Barcelona, Spain.}




\begin{abstract}

Spin Hall effects have surged as promising phenomena for spin logics operations without ferromagnets. However, the magnitude of the detected electric signals at room temperature in metallic systems has been so far underwhelming. Here, we demonstrate a two-order of magnitude enhancement of the signal in monolayer graphene/Pt devices when compared to their fully metallic counterparts. The enhancement stems in part from efficient spin injection and the large resistivity of graphene but we also observe 100\% spin absorption in Pt and find an unusually large effective spin Hall angle of up to 0.15. The large spin-to-charge conversion allows us to characterise spin precession in graphene under the presence of a magnetic field. Furthermore, by developing an analytical model based on the 1D diffusive spin-transport, we demonstrate that the effective spin-relaxation time in graphene can be accurately determined using the (inverse) spin Hall effect as a means of detection. This is a necessary step to gather full understanding of the consequences of spin absorption in spin Hall devices, which is known to suppress effective spin lifetimes in both metallic and graphene systems.
\end{abstract}


\maketitle

\section{Introduction}

Spin Hall effects (SHE) are a family of relativistic phenomena in which electrical currents generate transverse spin currents and vice versa as a consequence of spin-orbit coupling (SOC) \cite{r4}. In non-magnetic materials, they provide a versatile tool for the generation and detection of spin currents without the insertion of ferromagnets (FMs). Early theoretical investigations showed that the SHE can be of extrinsic \cite{r1, r2} or intrinsic \cite{r3, r4} origin. Experimental observations have been carried out in semiconductor and metallic systems by means of optical \cite{r5, r6}, electrical \cite{r7} and magnetization dynamics \cite{r8} methods, triggering many fundamental studies and potential applications \cite{r9}.

The exploration of the SHE using nonlocal (NL) measurements \cite{r7,SOV2007} have been frequently performed in metals using Pt as the large SOC material \cite{r10, r11}. In semiconductor systems, detection and modulation of the inverse SHE (ISHE) has been achieved in n-GaAs by means of spin precession in combination with external electric fields \cite{r12}. These studies have led to a better comprehension of the phenomena and more accurate determinations and control of the spin-to-charge conversion efficiency, which is quantified by the spin Hall angle ($\alpha_{SHE}$). For instance, $\alpha_{SHE}$ can be enhanced by using metallic alloys \cite{r13, r14}, or tuned by changing the electrical resistivity \cite{r15}. Moreover, recent experiments have demonstrated that spin currents generated by the SHE can be large enough to switch the magnetization of nanomagnets \cite{r16,r18} and induce fast domain wall motion \cite{r17}, highlighting the potential use of the SHE in conventional spintronic devices \cite{r18} and pointing the path towards future spintronics without FMs.

Additionally, 2D materials have emerged as promising system for spintronics owing to their tuneable electronic properties. In particular, graphene is attractive for both propagating and manipulating spin information over long distances because of its long spin-diffusion length $\lambda_s$ \cite{r19,SR2014, r20}. Studies of spin relaxation in graphene has been explored using standard Hanle measurements \cite{tombros, r22, r23, r24, r21, r26, volmer15} and out-of-plane precession to determine the spin relaxation anisotropy \cite{r27}. Improvements in the graphene quality have led to $\lambda_s$ in the range of tens of micrometers, which is already suitable for a number of applications \cite{kawakami15,r30}.

\begin{figure*}[ht]
\centering
\vspace{-10mm}\includegraphics[width=.8\paperwidth]{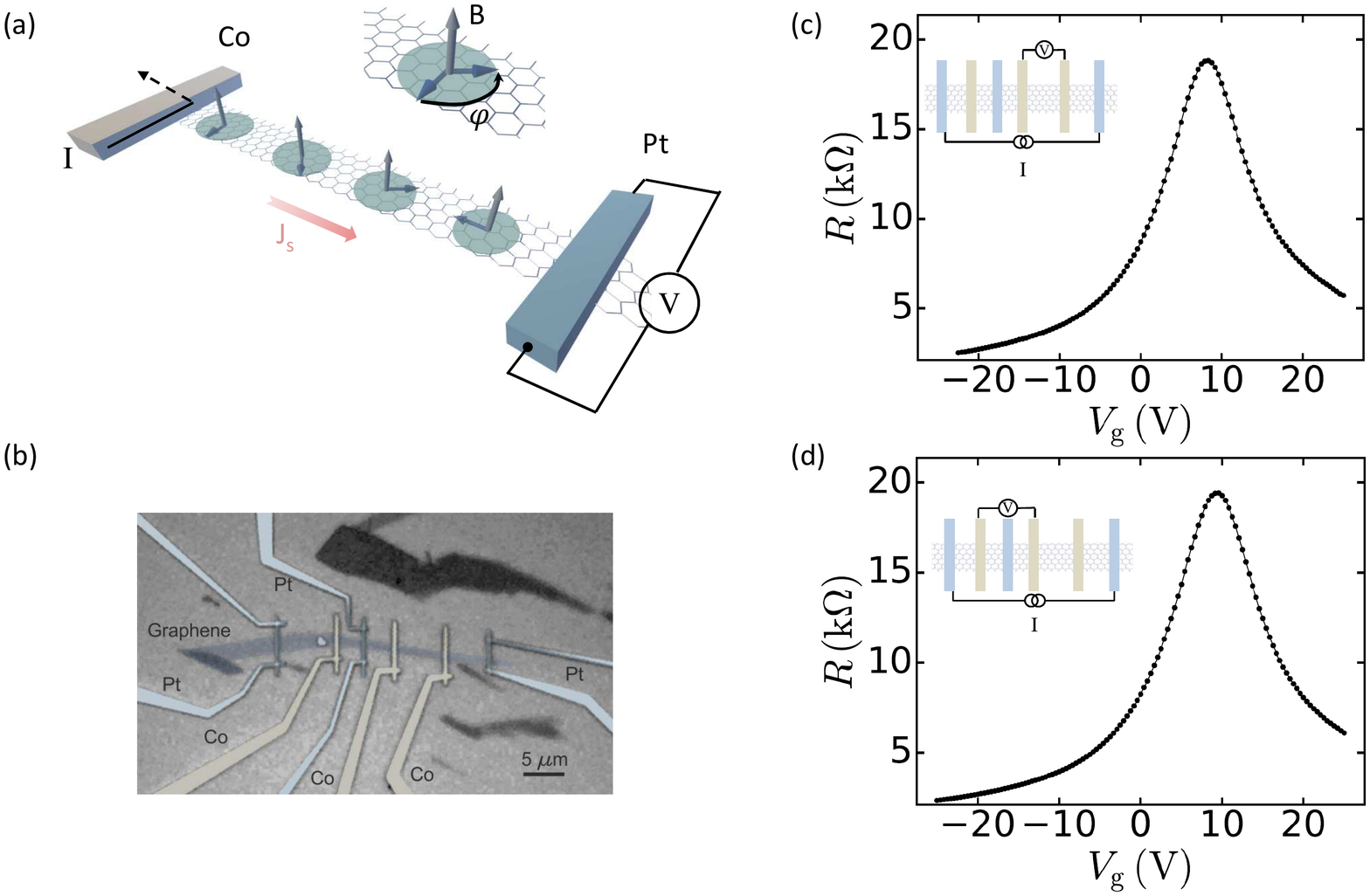}
\vspace{-15mm}
\caption{(a) Schematic representation of the probe configuration used to measure the inverse spin hall effect in our devices. The electrical current passing through the Co/graphene interface injects a spin current into the graphene channel. The spins are oriented along the Co magnetization (blue arrows). During the diffusion along the channel (red arrow), spins precess due to the action of the external magnetic field $B$ (gray arrow). The absorption into the graphene/Pt interface induces the resulting voltage due the ISHE in Pt. (b) False-colour optical image of a typical device. The two Co wires (150 and 200 nm wide) and two Pt wires (150 nm wide) located at the right side constitute the reference device for the standard Hanle measurements. The Pt wires in the middle and in the extreme right are used to perform the spin Hall measurements. (c,d) Probe configurations (insets) and corresponding four point resistances $R$ versus backgate voltage $V_g$ in the channel regions where standard Hanle and SHE measurements are carried-out.
}
\label{fig:Fig1}
\end{figure*}

Nonlocal signals detected in H-shape graphene structures decorated with metallic particles have been interpreted as due to large spin Hall effects, owing to SOC by proximity effects \cite{r31}. Theoretical analysis suggests that the proximity SOC should indeed enhance the spin Hall effect \cite{ferreira2014,r32}, but also that the signal in the H-shape structures can be affected by additional non spin-related contributions, which can lead to incorrect interpretations \cite{r32,axel2009}. More recent experiments using a ferromagnetic source/detector in multilayer graphene/Pt spin devices demonstrated large spin signals at room temperature, showing the potential of graphene-based devices for spin to charge conversion \cite{r33}.

Here, we unambiguously demonstrate the observation of the (I)SHE in monolayer graphene/Pt structures. We show that the spin-dependent properties of graphene can be fully characterised by combining spin precession experiments with the ISHE. We use an established spin-sensitive method \cite{r7} in which a spin polarised current is injected into the graphene channel by using a FM (Fig. 1a). The spin current precessing along the channel by the action of a magnetic field is then probed by measuring the resulting voltage generated by the ISHE in the Pt detector. Remarkably, the spin signals at room temperature are two orders of magnitude larger than those observed in equivalent devices with metallic channels. The spin current absorption by Pt becomes complete when the graphene is used in monolayer form, which has never been observed before. In addition, the use of highly-doped Si substrates allows us to modulate such signals by the action of a backgate voltage, providing a new tool to explore the spin relaxation dependence on the carrier density. In order to interpret the results, we develop a model to extract the spin Hall angle of Pt together with the graphene $\lambda_s$, and associated spin relaxation time $\tau_s$, in the region where the Pt wire is inserted. We find that $\tau_s$ is typically in the range of hundreds of picoseconds, in agreement with the estimations with standard Hanle measurements \cite{r19}. Our results therefore demonstrate a powerful method to study and characterise spin-dependent properties in non-magnetic materials, while the large signals open the door for room-temperature spintronic devices without the use of FMs.

\section{Device Fabrication}

\begin{figure*}[ht]
\centering
\vspace{-10mm} \makebox[\textwidth]{\includegraphics[width=.8\paperwidth]{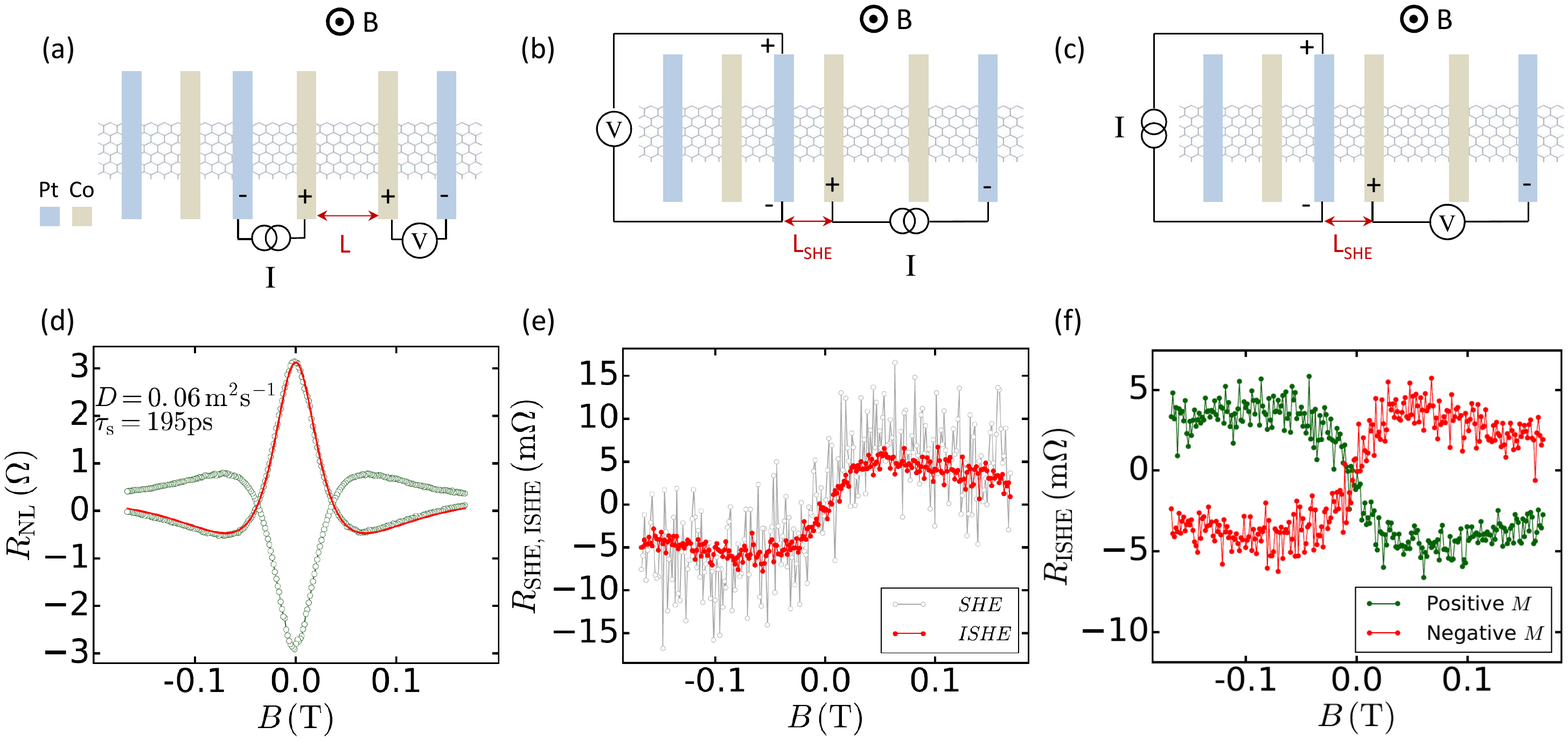}}
\vspace{-28mm}
\caption{(a)  Probe configuration and (d) standard Hanle measurement obtained for the parallel and antiparallel magnetic state of the device. The red curve shows the best fitting of the data from which the spin dependent parameters of graphene are extracted. (b,c) Probe configurations and (e) corresponding experimental results obtained for the SHE (gray curve) and ISHE (red and green curves) at $V_{g}=0$ in a device for which the channel length $L_{SHE}$ is 4 $\mu$m. The ISHE measurements in (f) show the dependance of the signal on the orientation of the magnetization $M$ of the injector. For clarity, small background signals have been extracted from the experimental results in (e) (8.5 m$\Omega$) and (f) (13.5 m$\Omega$).}
\label{fig:Fig2}
\end{figure*}

Figure 1b shows an optical micrograph of a typical device, which comprises Co and Pt electrodes connected to a graphene flake that is used as a spin channel. Graphene flakes are mechanically exfoliated from a highly oriented pyrolytic graphite source onto a p-doped Si substrate covered with 440-nm thick SiO$_{2}$, which is used to control the graphene carrier density. The Pt and FM Co electrodes are  defined in subsequent e-beam lithography steps and deposition in a high vacuum e-beam evaporator ($10^{-8}$ Torr). A 0.3 nm thick sticking layer of Ti is deposited prior to the deposition of Pt (12 nm thick; 150 nm wide). To assure good spin injection onto graphene 0.8 nm thick, TiO$_{2}$ barriers are fabricated in a controlled oxygen environment. The 30 nm thick FM electrodes are designed with widths of 150 and 200 nm in order to vary their coercive fields and enable access to the parallel and antiparallel magnetization configurations. After fabrication the samples are inspected with optical microscopy prior to the measurements and then fully characterized with SEM after concluding the experiments. The selected graphene flakes are typically $\sim 1$ $\mu$m wide and homogeneous along the device length.

\section{Results and discussion}

Transport measurements presented hereafter have been carried out at room temperature in a vacuum environment ($ 10^{-4}$ Torr). The injected current $I$ varies from 1 to 5 $\mu$A. Figures 1c and d display the resistance $R$ as a function of gate voltage $V_g$ at two different regions of the channel (the electrode configurations are shown in the insets). The first region (Fig. 1c), corresponding to the section of the channel where there is no Pt wire, is used as a reference to carry out standard Hanle measurements. The second region (Fig. 1d) corresponds to the zone where the Pt wire has been inserted in order to measure the SHE and characterize the spin absorption. It is observed that $R$ is not significantly modified by the presence of the Pt wire. In both regions graphene is weakly p-doped, with the charge neutrality point (CNP) located at $V_{g} \approx 10$ V. The  extracted mobility $\mu$ and residual carrier density $n_{0}$ are $\mu \approx  9000$ cm$^2$(Vs)$^{-1}$ and $n_0 \approx 2 \times 10^{11}$ cm$^{-2}$, which are comparable to those obtained in devices fabricated using angle evaporation without breaking vacuum \cite{r27}, demonstrating the reproducibility and quality of the device fabrication.

The spin transport is characterised by using standard non-local Hanle measurements. Here, an electrical current $I$ passing through the interface between the FM electrode and the graphene creates an out-of-equilibrium spin accumulation. Spin diffusion leads to the generation of a spin current along the channel, which is then detected by measuring the voltage $V_{NL}$ at the second FM electrode (see Fig. 2a). An external magnetic field $B$ applied out of the plane of the sample induces precession therefore modulating the detected signal. Figure 2d shows $R_{NL} = V_{NL}/I$ obtained for the parallel and antiparallel magnetic configuration at $V_{g}$= -20 V. The spin signal is recorded  while sweeping the magnetic field down. The long channel length of 7 $\mu$m, and the large Co/TiO$_2$ interface resistance, typically between 12 k$\Omega$ to 20 k$\Omega$, reduce the influence of the electrodes in the spin dynamics and help minimize the effects associated to the magneto resistance of graphene, as the magnetic field $B$ required for the complete characterisation of spin precession is small \cite{r27}. As observed in Fig. 2d, average $90^{\circ}$ and $180^{\circ}$ spin precession angles are achieved at relatively low $|B|\sim 40$ mT (zero crossing) and $|B|\sim 60$ mT (minima), respectively.

The output voltage $V_{NL}$ under the influence of $B$ can be modeled considering the diffusion of carriers in one dimension by \cite{tombros},

\begin{equation}
\label{eq:Hanle}
V_{NL}= \pm\frac{I p^{2}}{e^{2}NW_{G}}\int^{\infty}_{0}P(t)\cos(\omega_{L}t)\exp{(-\frac{t}{\tau_{s}})}dt
\end{equation}
where $p$, $e$, $N$ and $W_{G}$ are the effective spin polarization of the FM electrodes, the electron charge, the density of states and the width of graphene, respectively. $P(t)$ is the distribution function for electrons at the injector to reach the detector in time $t$ while $\exp{(-\frac{t}{\tau_{s}})}$ describes the spin relaxation in the channel. $P(t)$ is equal to $P(t)=\sqrt{1/4\pi Dt}.\exp(-L^2/4Dt)$, with $D$ the spin diffusion constant and $L$ the distance between injector and detector. The contribution of a carrier to $V_{NL}$ is proportional to the projection of the spin along the direction of the electrode magnetization, $\cos(\omega_{L}t)$ with $\omega_{L}$ the Larmor frequency. As observed in Fig. 2d, the experimental results are in good agreement with Eq. (1) (red curve), allowing us to extract the graphene spin dependent parameters, $\lambda_s\approx 3.4$ $\mu$m, and $D\approx 0.06$ m$^{2}$s$^{-1}$, which compare well with typical values obtained in graphene-based nonlocal devices on a Si/SiO$_2$ substrate \cite{r19}. Assuming equal polarization at both interfaces, the effective spin polarisation of the electrodes is $p = 0.16$.

Having characterised the devices with conventional Hanle experiments we carry out the measurements based on the ISHE using the probe configuration shown in Fig. 2b. A current $I$ is injected at a FM electrode, which generates a spin current $\mathbf{j}_s$ flowing towards Pt. As in the previous measurements, the external magnetic field is applied perpendicular to the graphene plane to induce spin precession. The spin current $\mathbf{j}_s$ is then absorbed at the Pt/graphene interface, resulting in a voltage $V_{ISHE}$, and the corresponding spin Hall resistance $R_{ISHE}=V_{ISHE}/I$, due to the ISHE in Pt (red curve in Fig. 2e). We observe that $R_{ISHE}$ is antisymmetric with $B$. At low $B$, $R_{ISHE}$ is approximately linear, whereas at higher $B$, $|V_{ISHE}|$ reaches a maximum and then decreases. Such a behavior is expected from a spin-related signal. Indeed, $V_{ISHE}$ is proportional to the cross product ($\mathbf{j}_{s}\times \mathbf{s}$) between the spin current $\mathbf{j}_s$ and the spin orientation $\mathbf{s}$ \cite{r2, r4}. Because the magnetization of the FM electrodes is oriented along their length, the spins at $B=0$ are parallel to the Pt wire and $V_{ISHE}$ is zero. Only when a non-zero magnetic field $B$ is applied, the spins are detected, as spin precession generates a spin component perpendicular to the Pt wire. When $B$ increases, $|V_{ISHE}|$ reaches a maximum and then decreases as the average spin precession angle exceeds $\sim 90^{\circ}$. The decrease of the signal at large $B$ ($> 100$ mT) is associated to spin dephasing. In contrast to the standard SHE detection \cite{r7, r10, r11, r13, r14, r15}, the FM electrodes do not rotate as the magnetic field is swept, therefore the signal is not saturated to a finite value and approaches zero instead.

As discussed previously, in the ISHE measurement the spin current injected into graphene is detected by measuring the resulting $V_{ISHE}$, and associated $R_{ISHE}$, in the Pt wire. In the direct SHE, the spin current is injected into graphene by passing an electrical current $I$ in the Pt wire. It is then detected by measuring the voltage $V_{SHE}$ between the FM electrode and the Pt wire located far from the injection point, from which the spin Hall resistance $R_{SHE}=V_{SHE}/I$ is obtained. Figure 2e demonstrates that $R_{SHE}=R_{ISHE}$, as expected from the Onsager symmetry relations \cite{r4,r7,SOV2007,r10}.

An additional confirmation of the spin-related origin of the signal is shown in Fig. 2f. Here, $R_{ISHE}$ is recorded before and after the magnetization of the injector is reversed with an in-plane magnetic field. Because the spin Hall voltage is proportional to $\mathbf{j}_{s}\times \mathbf{s}$, changing the orientation of the magnetization of the injector, and thus of the injected spins, directly modifies the detected voltage. As expected from a signal generated by spin currents, reversing the magnetization results in an inverted $R_{ISHE}$.
\begin{figure}[ht]
\centering
\includegraphics[width= 6.5 cm,height= 5.0 cm]{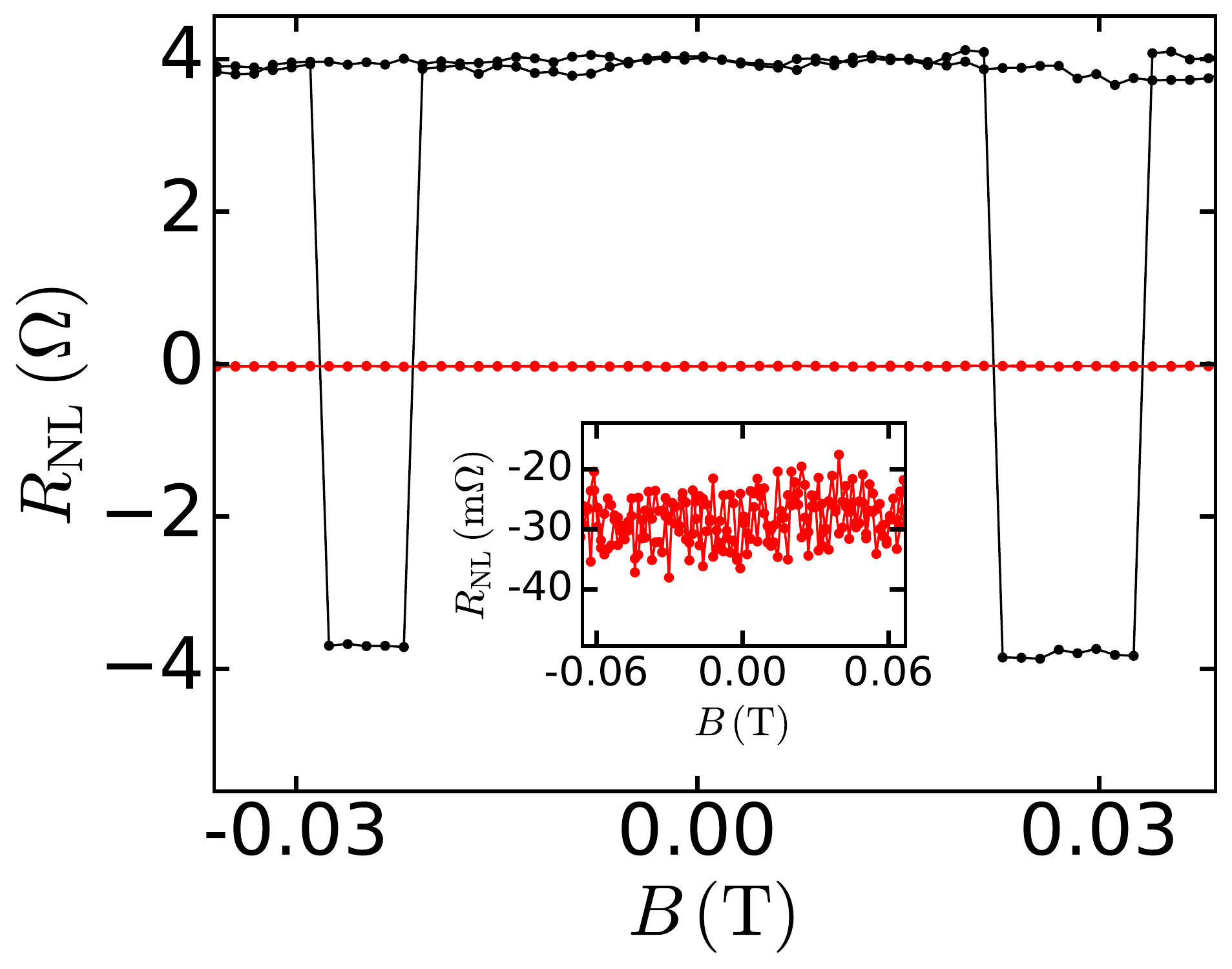}
\caption{Spin valve measurements performed in the device shown in Figure 1 at V$_{g}$=0 in the zones with and without the Pt wire. The channel lengths are L=7$\mu$m (black) and L=8$\mu$m (red). The black curve is the spin valve signal ($\approx$ 8m$\Omega$), measured in the zone without the Pt wire. The sharp variation in the resistance corresponds to the magnetization switching of the ferromagnetic electrodes. The spin valve measurement performed in the zone with the Pt wire (red curve), displays a flat signal with an upper amplitude of $\approx$10 m$\Omega$ (see inset), indicating that nearly all the current is absorbed by the Pt wire. Verification of good performance of all the interfaces is done by measuring the ISHE signal with the ferromagnetic contacts used for injection and detection in the spin valve measurements. }
\label{fig:Fig3}
\end{figure}

\begin{figure*}[ht]
\centering
\vspace{-30mm}\makebox[\textwidth]{\includegraphics[width=.75\paperwidth]{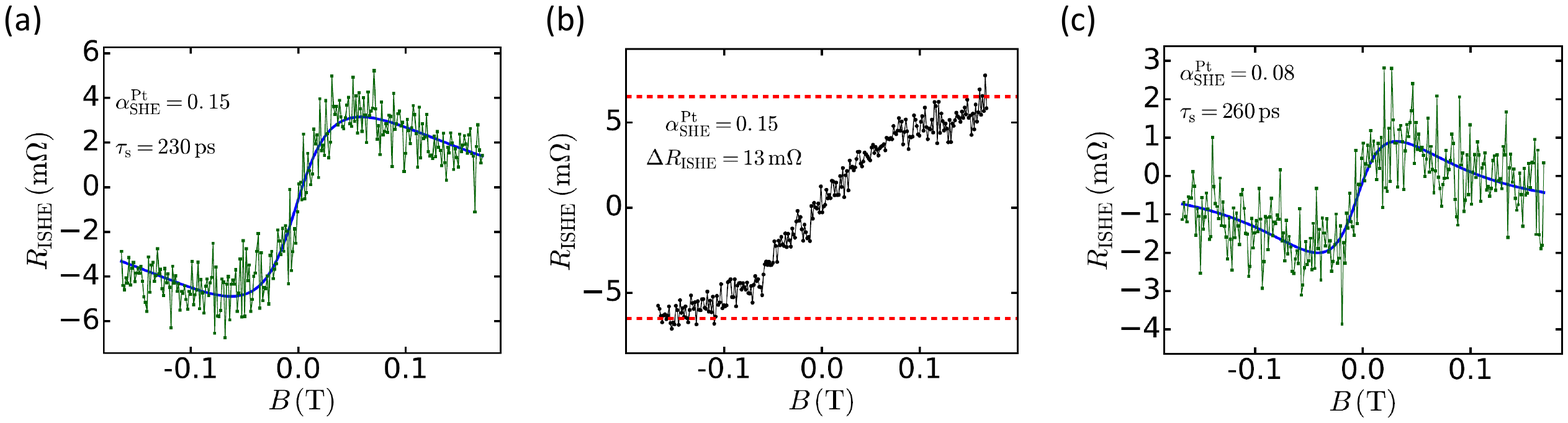}}
\vspace{-45mm}
\caption{Inverse spin Hall measurements performed at $V_{g} =-20$ V in devices with channel length $L_{SHE}=4$ $\mu$m (a,b) and $L_{SHE}=6.1$ $\mu$m (c), using the precessional method proposed in this work (a,c) and the standard rotation of the injector magnetization (b). The spin lifetime $\tau_s$ and spin Hall angle $\alpha^{Pt}_{SHE}$ are extracted by fitting the data to a spin-diffusion model described in the main text (blue curve). The dashed red lines in (b) indicate the saturation of the spin Hall signal, whose amplitude is $\sim 13$ m$\Omega$. A background signal of 13.5 m$\Omega$ has been extracted from the experimental results in (a) and (b).
}
\label{fig:Fig4}
\end{figure*}

Remarkably, the signal magnitude in our devices based on monolayer graphene can exceed 10 m$\Omega$, which is two orders of magnitude larger than the maximum reported value in metallic systems, even at cryogenic temperatures \cite{r13, r14, r15}. Such an increase is crucial to fully characterise the spin precession and determine the spin-dependent parameters in these type of structures at room temperature. The enhancement in comparison with metallic systems originates from the large spin injection achieved with resistive barriers and the reduction of the current shunting through the channel. Indeed, SHE detection using tunnel barriers has only been implemented using Al as a spin channel \cite{r7}, where the high spin polarisation ($p=0.28$) and efficient spin injection  permitted the determination of the spin Hall angle of Al [$\alpha^{Al}_{SHE}\approx (1-3)\times 10^{-4}$], which is two to three order of magnitude smaller than typical values reported for Pt \cite{r4}. Direct verification of the improved spin injection in our system has been done by calculating the spin current density at the injection point and comparing it with metallic systems. The current shunting through the channel is strongly reduced because the spin resistance of graphene [$\lambda_{G}R_{sq}(W_{G})^{-1}\approx 4.2$ k$\Omega$] is much larger than the one of platinum [$\rho_{Pt}\lambda_{Pt} (W_{Pt}W_{G})^{-1}\approx 0.01$ $\Omega$]. This feature can be important for an accurate extraction of the spin Hall angle since here all the current contributes to the ISHE signal, which is not the case for metallic systems. Note that the distance $L_{SHE}$ from the Hall cross to the injector is chosen so that significant precession is achieved at moderate $B$; larger $R_{SHE}$ by a factor two or more can easily be achieved by reducing such a distance, as it is done in metallic systems.

\begin{figure*}[ht]
\centering
\vspace{-15mm}\makebox[\textwidth]{\includegraphics[width=.75\paperwidth]{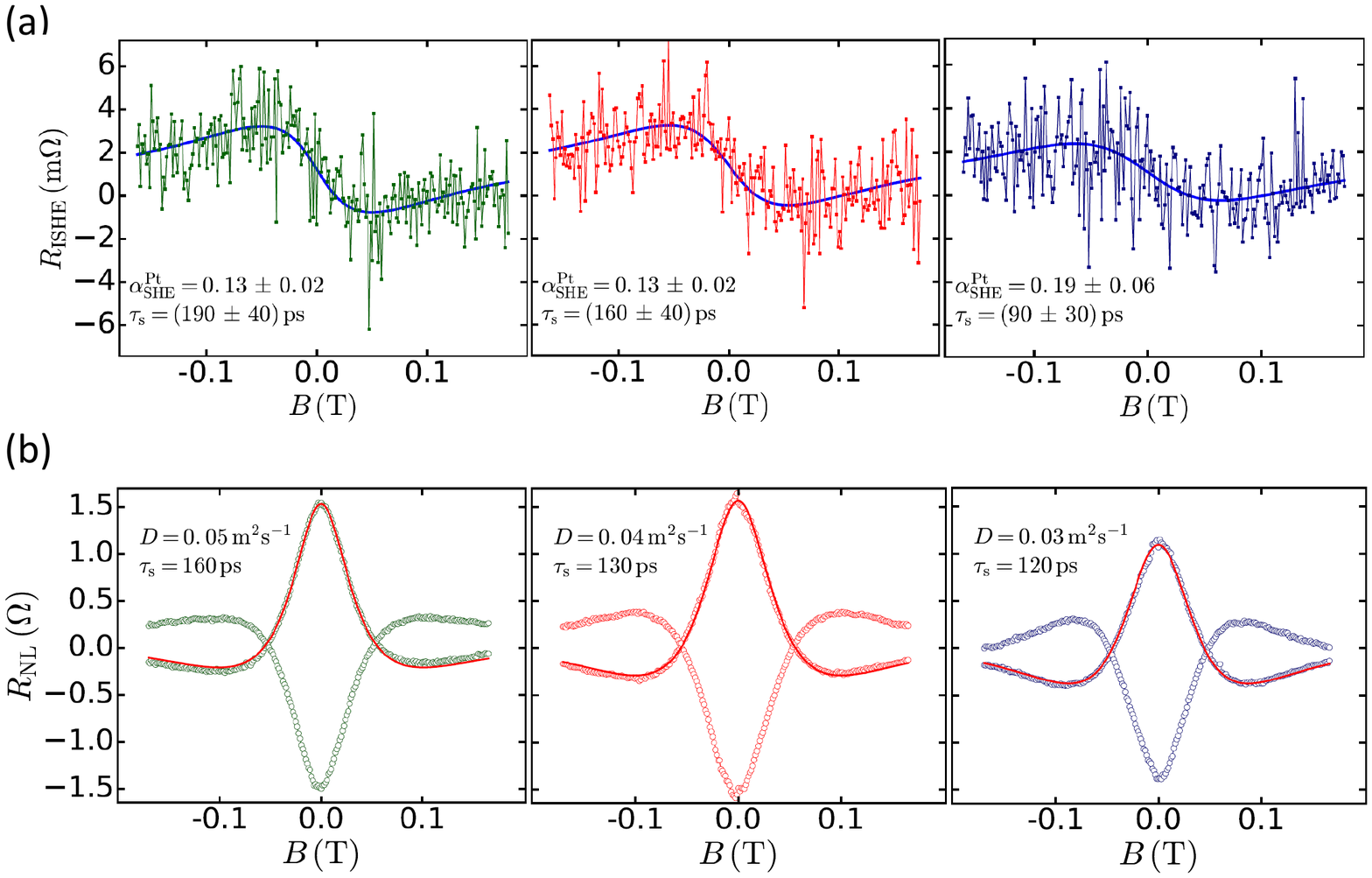}}
\vspace{-25mm}
\caption{Gate dependance of the SHE signal and comparison with standard Hanle measurements for device three. (a) Spin Hall measurements at gate voltages $V_{g}=-15$ V(green), -12 V (red) and -2 V (blue).(b) Hanle measurements obtained for $V_{g}=-15$ V(green), -5 V (red) and +0.5 V (blue). The gate voltage in the left and right panels represent measurements away from ($V_{g}=-15$ V) and near ($V_{g}=-2$ V) the charge neutrality point, respectively. $L_{SHE}= 5$ $\mu$m. For clarity, uncertainty in the extraction of the spin dependent parameters has not been inserted in the standard Hanle curves.
}
\label{fig:Fig5}
\end{figure*}

In order to quantitatively describe the results in Fig. 2, we develop a model based on one-dimensional diffusion equations in analogy to the model used for spin accumulation at a remote FM detector (Eq. \ref{eq:Hanle}). Assuming that the spin current in Pt flows perpendicular to the Pt/graphene interface, the resulting $V_{SHE}$ can be expressed as:

\begin{widetext}
\begin{equation}
\label{eq:Hall}
V_{SHE}= \frac{pI\rho L_{SHE} \lambda_{Pt} \alpha^{Pt}_{SHE}}{W_{Pt}t_{Pt}}\left[\frac{1-\exp(-\frac{t_{Pt}}{\lambda_{Pt}})}{1+\exp(-\frac{t_{Pt}}{\lambda_{Pt}})}\right]\int^{\infty}_{0}\frac{P(t)}{t}\sin(\omega_{L}t)\exp{(-\frac{t}{\tau_{s}})}dt
\end{equation}
\end{widetext}
where $\rho$, $\lambda_{Pt}$, $t_{Pt}$, $W_{Pt}$ and $\alpha^{Pt}_{SHE}$ are the electrical resistivity, spin relaxation length, thickness, width and spin Hall angle of Pt, respectively. Equation \ref{eq:Hall} is derived by assuming complete absorption of the spin current in the Pt wire, as described in the supplementary information. This assumption is supported by spin absorption experiments by measuring the spin signal in a nonlocal spin device with the Pt wire inserted between the spin injection and detection electrodes. As observed in Fig. 3, we find a completely flat $R_{NL}$ versus in plane magnetic field. As a comparison, the spin signal of a reference device with no inserted Pt wire is also shown. Considering the noise level as an upper bound for the change in $R_{NL}$ in the zone where Pt is inserted, its values is  $\Delta R_{NL} \sim 10$ m$\Omega$. It is almost three orders of magnitude smaller than the value found in the reference device, $\Delta R_{NL} \sim 8$ $\Omega$.

Equation \ref{eq:Hall} mainly differs from the standard Hanle result, Eq. \ref{eq:Hanle}, in the sinusoidal dependence on $B$ in the integrand, which originates from the $\mathbf{j}_{s}\times \mathbf{s}$ cross product. The sine factor is responsible of the antisymmetric behaviour of $V_{SHE}$ versus $B$ observed in the experiments, as well as the extrema at the average precession of $\sim 90^{\circ}$.

The spin relaxation time of graphene and the spin Hall angle of Pt can now be determined by fitting the experimental data using Eq. 2. To this end, the spin diffusion constant $D$ and effective spin polarisation values $p$ are extracted from the standard Hanle analysis in the same device (shown in Fig. 2d). The values of $\rho=46$ $\mu\Omega$cm, $t_{Pt} = 12$ nm, $W_{Pt}= 150$ nm and $L_{SHE}= 4$ $\mu$m are all measured, whereas only $\lambda_{Pt} = 5$ nm is taken from the literature, roughly corresponding to the average value at room temperature for Pt with similar resistivity \cite{r4}. With these assumptions, we obtain $\tau_{s} =(230\pm 25)$ ps, which is in good agreement with the value from the standard Hanle analysis, $\tau_{s}=(195\pm 5)$ ps. Because the spin precession lineshape is identical to that when spin absorption is not taken into account, this verifies that the effective $\tau_s$ is not significantly affected by spin absorption \cite{r23} (see supplementary information).

Interestingly, $\alpha^{Pt}_{SHE} =0.150\pm0.005$ is significantly larger than the largest values reported in metallic systems with comparable $\rho$,  $\alpha^{Pt}_{SHE}=0.085$, and close to the assumed values in multilayer graphene/Pt structures $\alpha^{Pt}_{SHE}= 0.234 \pm 0.025$ \cite{r33}, which were not measured but extrapolated from the results obtained in low resistivity Pt \cite{r15}. However, the resistivity of Pt in our device is significantly smaller than that in \cite{r33}, $\rho=134$ $\mu\Omega$cm, and thus the large $\alpha^{Pt}_{SHE}$ is unexpected \cite{r15}.

In order to go beyond this analysis and confirm our results, we perform complementary experiments using the standard SHE detection \cite{r7,r11} and investigate the carrier density dependence of $\alpha^{Pt}_{SHE}$ in additional devices. The standard SHE detection consists of using the probe configuration shown in Fig. 2b with an in-plane magnetic field along the channel axis. In this configuration, $R_{ISHE}$ follows a linear dependence at low magnetic field, which results from the dependence $\sin (\theta)$ on the rotation angle $\theta$ of the FM towards $B$ \cite{r7}. At large enough $B$, $R_{ISHE}$ saturates when the magnetization rotates $90^{\circ}$ and is parallel to $B$. The results are shown in Fig. 4b and demonstrate that saturation is reached at $\gtrsim 170$ mT. The extracted spin Hall angle with this approach is $\alpha^{Pt}_{SHE} = 0.15 \pm 0.01$, which is in excellent agreement with the previous analysis.

Figures 4c and 5 display $R_{ISHE}$ vs $B$ in devices two and three respectively. Device two is fabricated in a different region of the graphene flake used for the device in the prior discussion. Experimental results are shown in Fig. 4c. The magnitude of $R_{SHE}$ is smaller than in, for example, Fig. 4a because of the longer graphene channel $L_{SHE}=6.1$ $\mu$m. Using Eq. 2 results in $\tau_{s}  =(260\pm 50)$ ps, which is in good agreement with the values obtained from the standard Hanle analysis $\tau_{s}  =(195\pm 5)$ ps.  The extracted spin Hall angle is $\alpha^{Pt}_{SHE} =(0.08\pm 0.01)$.
Figure 5 shows the spin Hall and corresponding Hanle measurements for a third device, where $L_{SHE}=5$ $\mu$m. The measurements are carried out at the indicated gate voltages to show their carrier-density dependence. $R_{SHE}$ is observed to reach a minimum at the CNP, as commonly found in standard Hanle measurements. The diminution of $R_{SHE}$ as V$_{g}$ approaches the CNP is due to the decrease in $\tau_s$ combined with the conductivity mismatch effect expected for pinhole-like barriers and transparent interfaces, where the signal scales with the conductivity of the channel \cite{r23}. Using Eqs. 1 and 2 for the Hanle and spin Hall measurements, respectively, confirms the equivalence between the approaches to obtain $\tau_s$. The extracted spin Hall angles are $\alpha^{Pt}_{SHE}=(0.13\pm 0.02)$ for $V_g=-15$ V, $\alpha^{Pt}_{SHE} =(0.13 \pm 0.02)$ for $V_g=-12$ V and $\alpha^{Pt}_{SHE} =(0.19 \pm 0.06)$ for $V_g=-2$ V, which within the uncertainty of our measurements is gate independent. The estimated $\alpha^{Pt}_{SHE}$ obtained in three different devices is therefore in the range of 0.08-0.15. 

To calculate $\alpha^{Pt}_{SHE}$, we have taken the geometric mean of the polarization of two ferromagnetic electrodes to estimate $p$, and have assumed $\lambda_{Pt}= 5$ nm, which is roughly the average of the values found in the literature \cite{r4}. If $\lambda_{Pt}$ was larger than 5 nm, $\alpha^{Pt}_{SHE}$ would be somewhat overestimated by a few percentage points. However, recent experiments report $\lambda_{Pt}$ values that are lower than 5 nm for the resistivity of our Pt films \cite{r15}. If we take $\lambda_{Pt}$ = 1 nm, $\alpha^{Pt}_{SHE}$ would be underestimated by more than a factor 2. A larger $p$ of the injector electrode would lead to an overestimated $\alpha^{Pt}_{SHE}$, however, this is unlikely because $p$ in all our devices consistently lies between 0.12-0.16. Furthermore, 5 samples have been measured and from all of them we extract a large $\alpha^{Pt}_{SHE}$. In addition, our model considers that the spin current that arrives to Pt is fully absorbed by it. The lack of spin signal in the control spin absorption experiments (Fig. 3) validates this assumption. In any case, interface-driven spin relaxation or spin current flowing into the graphene towards the detector would make the actual spin current in Pt smaller, again underestimating $\alpha^{Pt}_{SHE}$. Therefore, our results suggest that $\alpha^{Pt}_{SHE}$ is larger than that observed in metallic systems with comparable Pt resistivities.

\section{Conclusions}

A two-order of magnitude enhancement of the spin Hall signal in graphene/Pt devices is observed when compared with their fully metallic counterparts. The enhanced signal allows us to characterise spin precession in graphene under the presence of a magnetic field. By developing an analytical model based on the 1D diffusive spin-transport, we demonstrate that the effective spin relaxation time in graphene can be accurately extracted using the (inverse) spin Hall effect as a means of detection. Good agreement is observed with the spin relaxation times obtained with standard Hanle measurements, based on the detection of spin accumulation away from Pt. This is relevant since a change in the spin relaxation is typically not taken into account in the standard spin Hall experiments \cite{r11}, where the spin lifetimes are assumed from control experiments.

The enhanced spin Hall signal is due to the efficient spin injection into graphene in combination with a current shunting suppression and a consistently large spin-to-charge conversion, as reflected in the large spin Hall angle. In multilayer graphene /Pt devices, a similar enhancement was attributed to the suppressed shunting and the long spin relaxation length of multilayer graphene, relative to that in metallic channels \cite{r33}. However, the change in spin relaxation length can be overcome with suitable device design by simply adapting the spin channel length $L_{SHE}$. Indeed, devices with metallic systems are designed to have $L_{SHE} \sim 0.5$ $\mu$m, which is of the same order of magnitude that, for example, the room-temperature spin relaxation length of Cu, a common material used for nonlocal spin Hall detection \cite{r4}. Because in our devices $\lambda_s < L_{SHE}$, the longer spin relaxation length of graphene $\lambda_s$ cannot play a central role. The shunting effect is very sensitive to geometrical dimensions. Typical values of the shunting reduction factor for the devices with the largest spin-to-charge conversion are about 0.2-0.3 \cite{r4,r14}, thus current shunting can account for a factor 3 to 5 when comparing with those devices. Finally, our experiments show that the spin Hall angle significantly exceeds those obtained in fully metallic systems, which could indicate the presence of interfacial spin-orbit effects between graphene and Pt that are not taken into account in our analysis. Indeed, a significant modification of the spin structure and an enhanced induced spin-orbit splitting of the graphene on Pt were recently reported \cite{r34}. However, further experiments are required to establish the presence of interfacial effects, in particular to address sample to sample variability associated to the effective polarization of the ferromagnetic electrodes and the dependence of the spin Hall angle with the Pt thickness.

\section{Acknowledgments}

This research was supported by the European Union's Horizon 2020 research and innovation programme under grant agreement No. 696656, by the European Research Council under Grant Agreement No. 308023 SPINBOUND, by the Spanish Ministry of Economy and Competitiveness, MINECO (under Contracts No. MAT2013-46785-P, No MAT2016-75952-R and Severo Ochoa No. SEV-2013-0295), and by
the CERCA Programme and the Secretariat for Universities and Research, Knowledge Department of the Generalitat de Catalunya 2014SGR56. J.F.S. acknowledges support from the MINECO Juan de la Cierva programme and F.B. and M.V.C from the MINECO Ram\'{o}n y Cajal programme.

\bibliography{refs_SHE}

\end{document}


\title{\textbf{Spin precession and the spin Hall effect in graphene/Pt nanostructures\\Supplementary Information}}

\author{W. Savero Torres}
\affiliation{Catalan Institute of Nanoscience and Nanotechnology (ICN2), CSIC and The Barcelona Institute of Science and Technology (BIST), Campus UAB, Bellaterra, 08193 Barcelona, Spain.}

\author{J.F. Sierra}
\affiliation{Catalan Institute of Nanoscience and Nanotechnology (ICN2), CSIC and The Barcelona Institute of Science and Technology (BIST), Campus UAB, Bellaterra, 08193 Barcelona, Spain.}

\author{L.A. Ben\'{i}tez}
\affiliation{Catalan Institute of Nanoscience and Nanotechnology (ICN2), CSIC and The Barcelona Institute of Science and Technology (BIST), Campus UAB, Bellaterra, 08193 Barcelona, Spain.}
\affiliation{Universidad Autonoma de Barcelona, Bellaterra, 08010 Barcelona, Spain}

\author{F. Bonell}
\affiliation{Catalan Institute of Nanoscience and Nanotechnology (ICN2), CSIC and The Barcelona Institute of Science and Technology (BIST), Campus UAB, Bellaterra, 08193 Barcelona, Spain.}

\author{M.V. Costache}
\affiliation{Catalan Institute of Nanoscience and Nanotechnology (ICN2), CSIC and The Barcelona Institute of Science and Technology (BIST), Campus UAB, Bellaterra, 08193 Barcelona, Spain.}

\author{S.O. Valenzuela}
\affiliation{Catalan Institute of Nanoscience and Nanotechnology (ICN2), CSIC and The Barcelona Institute of Science and Technology (BIST), Campus UAB, Bellaterra, 08193 Barcelona, Spain.}
\affiliation{Institucion Catalana de Recerca i Estudis Avan\c{c}ats (ICREA), 08070 Barcelona, Spain.}

\maketitle


In the following, the analytical expressions for the spin Hall voltage $V_{SHE}$ used in the main text are obtained. The standard procedure used in the literature is described and adapted to introduce spin injection with tunnel barriers. This procedure focuses on the non-precessing spin component and the rotation of the ferromagnetic injector magnetization \cite{SOV2006,SOV2007} combined with spin absorption in Pt \cite{kimura2007}. Afterwards, the model is extended to introduce spin precession. In all cases, we use a one-dimensional approximation.

Figure 1 shows the probe configuration and the geometry used to describe the relevant parameters for the calculation. The coordinate origin is defined at the position of the injector electrode. $L_{SHE}$ is the distance from the injector to the spin-Hall cross, while $W_{Pt}$, $w_{G}$, $t_{Pt}$ are the platinum width, graphene width and platinum thickness, respectively. $S=W_{Pt}*w_{G}$ is the area of the Pt/graphene contact, where the current is absorbed. In the standard measurement, the magnetic field $B$ is applied along $x$ direction, while in the spin precession measurements, $B$ is out of the sample plane, in the $z$ direction.

\begin{figure}[hb]

\begin{centering}
\includegraphics[width=9cm]{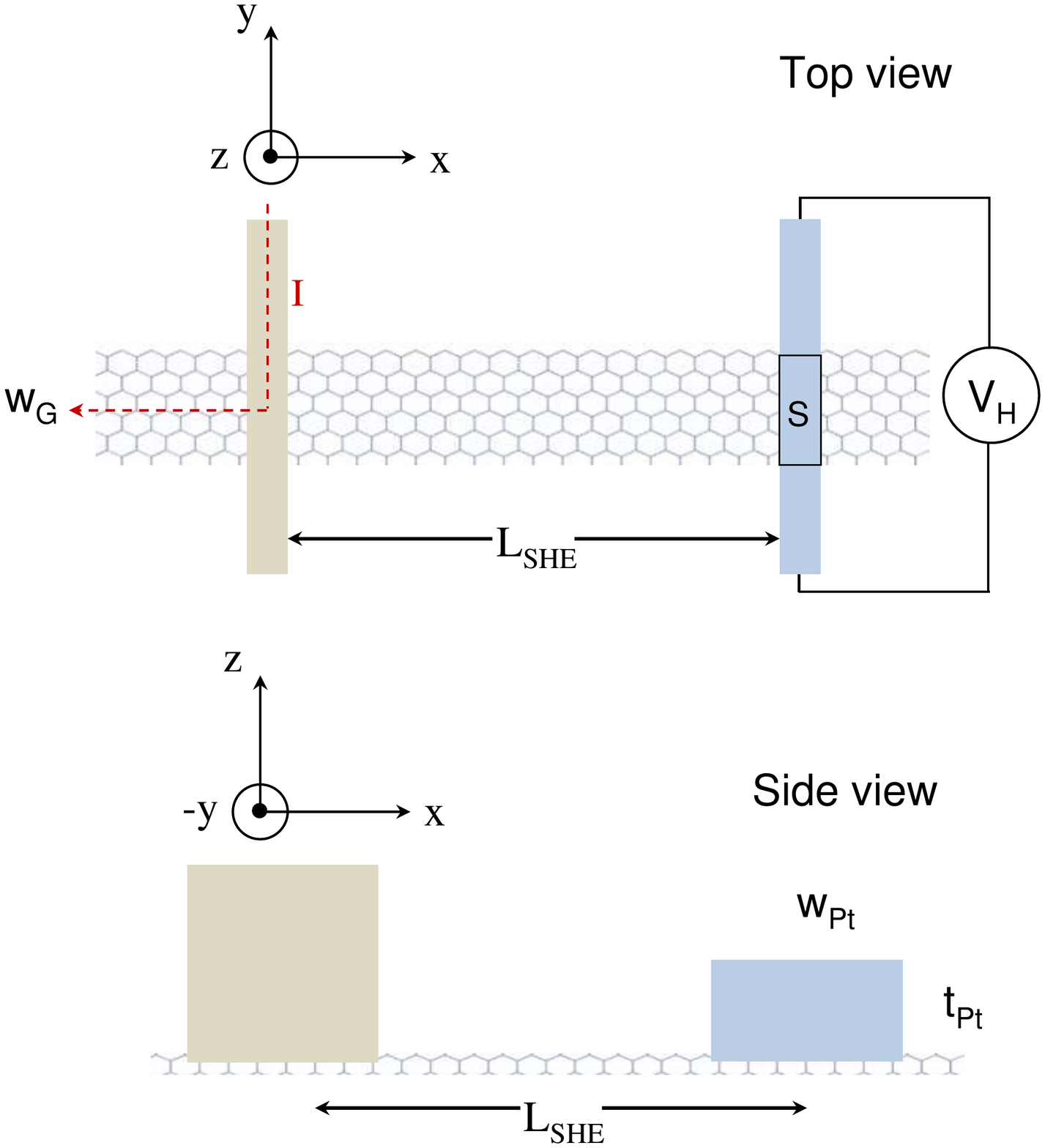}\caption{Schematic representation of the probe measurement\label{fig:a}}

\par\end{centering}

\end{figure}

The charge current $J^{T}_{q}$ in the Pt electrode at $B=0$ can be written as \cite{s0}:

$$  \textbf{J}^{T}_{q} = \alpha^{Pt}_{SHE}[\hat{\textbf{\i}} \times \textbf{J}_{s}] + \sigma_{Pt}\textbf{E}$$
where $\alpha^{Pt}_{SHE}$ and $\sigma_{Pt}$ are the Hall angle and the electrical conductivity of Pt respectively and $\textbf{J}_{s}$ is the spin current reaching the detector. The first term describes the spin to charge conversion due to the ISHE while the second term contains the electrical field that counteracts the charge current in the steady state,  in which both components compensate, therefore:
 $$\textbf{0} = \alpha^{Pt}_{SHE}[\hat{\textbf{\i}} \times \textbf{J}_{s}] + \sigma_{Pt}\textbf{E}$$

At $x=L_{SHE}$ the average spin current over the Pt wire is $\textbf{J}_{s}= \bar{J}_{s}\hat{\textbf{k}}$ and the resulting electric field along the $y$ direction is $\textbf{E}= E_{y}\hat{\textbf{\j}}$. The electric field $E_{y}$ is then :
$$E_{y}= \alpha^{Pt}_{SHE}\rho_{Pt}\bar{J}_{s}$$

The integration over the graphene width provides the Hall voltage $V_{SHE}$:
\begin{equation}
V_{SHE}= \alpha^{Pt}_{SHE}\rho_{Pt}w_{G}\bar{J}_{s}
\end{equation}

$\bar{J}_{s}$ is derived from the spin accumulation $\mu_{s}$ generated in the Pt wire, defined by $\mu^{Pt}_{s}=(\mu_{+}-\mu_{-})/2$, where $\mu_{\pm}$ are the electrochemical potentials for each spin population. Considering a 1D diffusive spin transport, $\mu^{Pt}_{s}$ is:
$$\mu^{Pt}_{s}(z)= A \exp(z/\lambda_{Pt})+B \exp(-z/\lambda_{Pt})$$
where $z$ indicates the dependence on the Pt thickness. Since $J_{s}=-\sigma/e\nabla\mu_{s}$, the spin current density $J^{Pt}_{s}$ can be expressed as:
$$J^{Pt}_{s}(z)= \frac{-\sigma_{Pt}}{e\lambda_{Pt}}[A \exp(z/\lambda_{Pt})-B \exp(-z/\lambda_{Pt})]$$

To determine the constants A and B, we consider the following boundary conditions: 1) At $z=0$, $J^{Pt}_{s}= J_{z0}$, where $J_{z0}$ is the current density reaching the detector after diffusing from the injection point and 2) At $z=t_{Pt}$, $J^{Pt}_{s}= 0$. From these conditions, we obtain:

$$B=\frac{-J_{z0}e\lambda_{Pt}}{\sigma_{Pt}}\frac{1}{\exp(-2t_{Pt}/\lambda_{Pt})-1}$$
$$A=\frac{-J_{z0}e\lambda_{Pt}}{\sigma_{Pt}}\frac{\exp(-2t_{Pt}/\lambda_{Pt})}{\exp(-2t_{Pt}/\lambda_{Pt})-1}$$

The spin current density along the thickness $J_{s}(z)$ is therefore given by:
\begin{widetext}
$$J^{Pt}_{s}(z)=J_{z0}[\frac{\exp(-2t_{Pt}/\lambda_{Pt})}{\exp(-2t_{Pt}/\lambda_{Pt})-1}]\exp(z/\lambda_{Pt})-J_{z0}[\frac{1}{\exp(-2t_{Pt}/\lambda_{Pt})-1}]\exp(-z/\lambda_{Pt})$$
\end{widetext}

\newpage

The average of $J^{Pt}_{s}(z)$, $\bar{J}_{s}$, is calculated from \cite{s1,s2, s3, s4}:

\begin{equation}
\bar{J}_{s}= \frac{1}{t_{Pt}}\int^{t_{Pt}}_{0}J^{Pt}_{s}(z)dz=\frac{J_{z0}\lambda_{Pt}}{t_{Pt}} \frac{[1-\exp(-t_{Pt}/\lambda_{Pt})]}{[1+\exp(-t_{Pt}/\lambda_{Pt})]}
\end{equation}

The current $J_{z0}$ at the interface between graphene and Pt is determined from the boundary conditions with the graphene region. It is straightforward to obtain the full solution analytically, however, a much simpler expression for $J_{z0}$ can be found by noting that $\mu \sim 0$ at the graphene/Pt interface due to the fast relaxation in Pt. Such a solution is an excellent approximation to the full solution for the relevant magnitudes of the involved parameters, and can be easily deduced as follows.

\subsubsection{Case 1: Standard spin Hall measurements}

We start by considering the solution of the diffusion equation for the graphene wire without Pt. Because the spin current in graphene $J_{s}\rightarrow 0$ when $x\rightarrow\infty$ and we use tunnel barriers for spin injection, $J_{z0}$ can be approximated by \cite{s0},

\begin{equation}
J_{z0}\approx \frac{pI}{2S}\exp(-L_{SHE}/\lambda_{G})
\end{equation}
where $\lambda_{G}$ is the spin relaxation length of graphene, $p$ is the effective spin polarization and $S$ the contact surface shown in Fig. 1. Note that in Eq. 3, it is assumed that the magnetization of the injector is saturated along the the magnetic field. If the rotation is not complete, the additional factor $\sin \gamma$ should be added to account for the magnetization tilting angle $\gamma$. \cite{SOV2007}.

Because the spin current is fully absorbed by the Pt wire, no spin current flows away from Pt in the graphene opposite to the side of the injector, where $\mu \sim 0$, as verified experimentally. In addition, this implies that the contact resistance is very small and does not play a significant role in determining $J_{z0}$. Considering that $\lambda_{G}\gg \lambda_{Pt}$ and that the conductivity of Pt is much larger than that of graphene, it can be easily verified that $J_{z0}$ is weakly dependent on the exact value of $\lambda_{Pt}$. Indeed, the region of graphene on the other side of the injector can safely be ignored. The full solution of the spin diffusion in our device is thus nearly identical to a problem involving two spin sources of opposite polarities separated by $2L_{SHE}$ without the Pt wire. In this equivalent problem, the condition $\mu = 0$ in the equidistant point to the sources (Pt wire position)is rigorously satisfied. 

The current $J_{z0}$ in the middle of the two sources, which is absorbed by Pt, is therefore twice the current in Eq. 3,

\begin{equation}
J_{z0}\approx \frac{pI}{S}\exp(-L_{SHE}/\lambda_{G})
\end{equation}

By replacing Eq. 2 and 4 in Eq. 1, the inverse spin Hall voltage for this case is:
\begin{equation}
\label{eq:Hall}
V_{SHE}= \frac{pI\alpha^{Pt}_{SHE}\rho_{Pt}\lambda_{Pt}}{w_{Pt}t_{Pt}} \frac{[1-\exp(-t_{Pt}/\lambda_{Pt})]}{[1+\exp(-t_{Pt}/\lambda_{Pt})]}\exp(-L_{SHE}/\lambda_{G})
\end{equation}

\subsubsection{Case 2: Precessional spin Hall measurement}
In this case, the magnetic field, applied out of the sample plane, induces spin current precession therefore changing the expression for $J_{z0}$. To determine $J_{z0}$, we first calculate the contribution from the precessing non-equilibrium spin density $n_{+}-n_{-}$ reaching the detector location, which is related to the electrochemical potentials $\mu_{\pm}$ as $\mu_{\pm}=\frac{n_{\pm}}{N_{\pm}(E_{F})}$, where $n_{\pm}$ and $N_{\pm}(E_{F})$ are the spin density and the density of states at the Fermi level for each spin orientation \cite{s2}.

An argument analogous to that in the previous case can be made. For the spins freely diffusing in graphene, without the Pt wire, the non equilibrium density of spins perpendicular to the spin Hall bar is:
$$n_{+}-n_{-}=\frac{Ip}{eS} \int^{\infty}_{0} \frac{\exp(-x^{2}/4Dt)}{\sqrt{4\pi Dt}}\sin(w_{L}t)\exp(-t/\tau_{s})dt$$
where $Ip/eS$ is the spin injection rate and $\frac{\exp(-x^{2}/4Dt)}{\sqrt{4\pi Dt}}\exp(-t/\tau_{s})$  the probability to reach the detector at time $t$, $w_{L}t$ is the precession angle, and $w_{L}$ the Larmor frequency. Since in the spin channel no charge current flows, $N_{+}(E_{F})+ N_{-}(E_{F}) = N(E_{F})$ and $N_{+}(E_{F})= N_{-}(E_{F}) = N(E_{F})/2$. The corresponding spin accumulation $\mu^{N}_{s}$ is then given by:
 $$\mu^{N}_{s}=\frac{Ip}{eSN(E_{F})} \int^{\infty}_{0}\frac{ \exp(-x^{2}/4Dt)}{\sqrt{4\pi Dt}}\sin(w_{L}t)\exp(-t/\tau_{s})dt$$
As $J^{N}_{s}=-\sigma_{N}/e\nabla\mu^{N}_{s}$, with $\sigma_{N}$ the conductivity of the channel, the spin current  is:
$$J^{N}_{s}=\frac{Ipx\sigma}{2De^{2}N(E_{F})S} \int^{\infty}_{0}\frac{\exp(-x^{2}/4Dt)}{t\sqrt{4\pi Dt}}\sin(w_{L}t)\exp(-t/\tau_{s})dt$$

Using the Einstein relation: $\sigma=e^{2}N(E_{F})D$, then:
$$J^{N}_{s}=\frac{Ipx}{2S} \int^{\infty}_{0}\frac{ \exp(-x^{2}/4Dt)}{t\sqrt{4\pi Dt}}\sin(w_{L}t)\exp(-t/\tau_{s})dt$$
At $x=L_{SHE}$ the spin current $J^{N}_{s}$= $J_{z0}$ is given by:
\begin{equation}
J_{z0}=\frac{IpL_{SHE}}{2S} \int^{\infty}_{0}\frac{ P(t)}{t}\sin(w_{L}t)\exp(-t/\tau_{s})dt
\end{equation}

By multiplying this expression by a factor 2, resulting from the condition $\mu=0$, and substituting in Eq. 1, we find the expression for the Hall voltage $V_{SHE}$ shown in Eq. 2 of the main text,
\begin{widetext}
\begin{equation}
V_{SHE}= \frac{I\alpha^{Pt}_{SHE}\rho_{Pt}\lambda_{Pt}pL_{SHE}}{W_{Pt}t_{Pt}} \frac{[1-\exp(-t_{Pt}/\lambda_{Pt})]}{[1+\exp(-t_{Pt}/\lambda_{Pt})]} \int^{\infty}_{0}\frac{ P(t)}{t}\sin(w_{L}t)\exp(-t/\tau_{s})dt
\end{equation}
\end{widetext}
where the precession of $J_{z0}$ inside the Pt wire is neglected because of the very small dephasing times in Pt($\tau_{s}\approx 3\times10^{-13}$s \cite{s6}).

The equivalence between the standard (Eq. 5) and precessional approach (Eq. 7) in the limit of no spin precession can be verified by substituting $\sin(w_{L}t)=1$ in the integral of Eq. 7 and noting that $\lambda_{G}=\sqrt{Dt}$.

\bibliography{refs_sup}